\documentclass[%
 reprint,
 amsmath,amssymb,
 aps,
]{revtex4-2}

\usepackage{graphicx}
\usepackage{dcolumn}
\usepackage{bm}
\usepackage[dvipsnames]{xcolor}
\usepackage{color}
\usepackage{soul}

\newcommand{\pp}[2]{ \frac{\partial #1}{\partial #2}}
\newcommand{\ljump}{{[\![}}
\newcommand{\rjump}{{]\!]}}

\newcommand{\mrm}[1]{\mathrm{#1}}
\newcommand{\bbm}[1]{\boldsymbol{\mrm{#1}}}
\newcommand{\bv}[0]{\bm{v}}

\definecolor{lightblue}{rgb}{0,0.58,0.71}
\newcommand{\zx}[1]{#1}
\newcommand{\zxy}[1]{#1}
\graphicspath{{./note_figures/}
{./}
{./revised_figures/}
}

\begin{document}

\preprint{APS/123-QED}

\title{Instability of a fluctuating biomimetic membrane driven by an applied uniform DC electric field}

\author{Zongxin Yu}
\affiliation{%
Engineering Sciences and Applied Mathematics, Northwestern University
}%

\author{Shuozhen Zhao}%
\affiliation{%
Engineering Sciences and Applied Mathematics, Northwestern University
}

\author{Michael J. Miksis}%
\affiliation{%
Engineering Sciences and Applied Mathematics, Northwestern University
}

\author{Petia M. Vlahovska}%
\affiliation{%
Engineering Sciences and Applied Mathematics, Northwestern University
}

\date{Aug 5 2025 }

\begin{abstract} 
The linear stability of a lipid membrane under a DC electric field, applied perpendicularly to the interface, is investigated in the electrokinetic framework, taking into account the dynamics of the Debye layers formed near the membrane. The perturbed charge in the Debye layers redistributes and generates destabilizing Maxwell stress on the membrane, which outweighs the stabilizing contribution from the electrical body force, leading to a net destabilizing effect. The instability is suppressed as the difference in the electrolyte concentration of the solutions separated by the membrane increases, due to a weakened base state electric field near the membrane. This result contrasts with the destabilizing effect predicted using the leaky dielectric model in cases of asymmetric conductivity. We attribute this difference to the varying assumptions about the perturbation amplitude relative to the Debye length, which result in different regimes of validity for the linear stability analysis within these two frameworks.
\end{abstract}

\maketitle
 \section{Introduction}


Cells are enveloped by thin membranes that isolate the cell interior from the environment.
The response of biological membranes and cells to electric fields has received a lot of attention, due to practical applications in cell manipulation, such as fusion and electroporation~\cite{Jordan_1989,Weaver96_bio}. These processes rely on the controlled destabilization of membranes~\cite{Teissie05_bio,Dimova09_sm}, where electric fields influence membrane fluctuations, alter geometry through stretching and compression~\cite{Riske05_bj,Riske06_bj}, and modify mechanical properties such as tension and bending~\cite{Sens02_prl}. Despite extensive experimental~\cite{Riske05_bj,Riske06_bj} and theoretical studies, the fundamental mechanisms governing membrane destabilization in electric fields remain an area of ongoing research. Theoretical models for instability analysis involve two main approaches: the leaky dielectric model (LDM) and the Poisson-Nernst-Planck (PNP) model.

The LDM assumes that free charge is confined to interfaces separating media with different electric properties (conductivity and permittivity), while the bulk region is charge-free. Pioneering work in this area dates back to the 1960s, with notable contributions from G. I. Taylor and J. R. Melcher~\cite{Taylor65_jfm,Melcher65_pof}, who investigated the deformation and stability of fluid-fluid interfaces under electric fields. The LDM was extended to model electrohydrodynamics of bio-membranes by treating the thin membrane as a zero-thickness interface with effective electric properties, e.g.,  capacitance $C_m$. Studies using this model have reported instabilities induced by various mechanisms, including ion currents across the membrane subjected to DC normal fields~\cite{Seiwert12_jfm} and tangent fields~\cite{Young14_fjm}, and capacitive current upon application of a DC field \cite{Schwalbe11_pof}, and in AC fields with periods shorter than the capacitor charging time~\cite{Seiwert13_pre}. 

 Although effective on a theoretical basis, the above studies neglect the thermal motion of the free charges. In a real system, the ions near an interface form a diffuse layer. This layer has a characteristic thickness, called the Debye screening length, which is typically a few nanometers. The \zxy{PNP} equations govern the dynamics of the charges within electrolytic materials, and have been used extensively in electrokinetics \zx{(EK)} to study membrane or fluid interfaces~\cite{Saville97_arfm}. Refs.~\cite{kumaran2000_prl,kumaran01_pre} investigated curvature-surface charge density coupling induced instabilities of a membrane.
 Refs.~\cite{Lomholt07_pre,Loubet13_pre} analyzed the electrostatic correction to the  tension and bending rigidity, and the instability of a finite-thickness membrane. Refs.~\cite{Lacoste09_epje,Ziebert10_pre} proposed a zero-thickness model with the Robin-type boundary condition, predicting electrokinetic corrections to the elastic moduli of the membrane, and a new destabilizing term in the growth rate of the membrane undulations due to ion current through the membrane. These studies report membrane instability arising from unbalanced electrical stress on the membrane, due to either surface charge~\cite{kumaran2000_prl,kumaran01_pre} or charge within the Debye layer~\cite{Lomholt07_pre,Loubet13_pre,Lacoste09_epje,Ziebert10_pre}. Similar instabilities have also been observed in fluid-fluid interface systems, such as the perfect-dielectric liquid/electrolyte interfaces~\cite{Thaokar14_pre}.

In a steady DC electric field normal to a non-conducting membrane, the LDM predicts that the fully-charged membrane capacitor suppresses instabilities~\cite{Seiwert12_jfm}. Instability arises only in AC fields~\cite{Seiwert13_pre} or during the startup of a DC field~\cite{Schwalbe11_pof}, where the membrane acts as a short-circuited capacitor, leading to charge imbalances on its surfaces. 
However, the PNP model predictis instability even under DC fields~\cite{Lacoste09_epje,Ziebert10_pre}. The physical mechanisms of this instability remain less intuitive and we provide an explanation by examining  the charge distribution in the Debye layers near the membrane. Refs.~\cite{Lacoste09_epje,Ziebert10_pre} focused on symmetric membranes, i.e., separating electrolyte solutions with the same ions and concentrations. \zx{However, in the experiments with vesicles in electric fields\cite{riske2005electro,aranda2008morphological,Salipante:2012,salipante2014vesicle,Salipante:2015}, the electrolyte concentrations inside and outside the vesicle are typically different, necessitating an analysis of asymmetric membranes.
We extend the analysis to account for this asymmetry and show that differences in ionic concentration can have a stabilizing effect, reducing the instability predicted by the symmetric model. Our analysis highlights the critical role of ionic distribution and membrane asymmetry in determining stability under electric fields.}

In Section~\ref{sec:formulation}, we introduce the governing equations for two electrolyte solutions of ions, separated by a thin ion-impermeable membrane. Section~\ref{sec:solution} discusses the base state solution and its linear stability. We explore the effects of key parameters on the instability in Section~\ref{sec: alpha 0}, identifying the sources of instability and elucidating why conductivity asymmetry reduces it. In addition, we show that asymmetric conductivity gives rise to the spontaneous curvature. Finally, conclusions and avenues for further research are summarized in Section~\ref{sec:conclusion}.
 \section{Problem formulation and Governing equations}
 \label{sec:formulation}
\begin{figure}[ht!]
    \centering
\includegraphics[keepaspectratio=true,width=0.8\columnwidth]{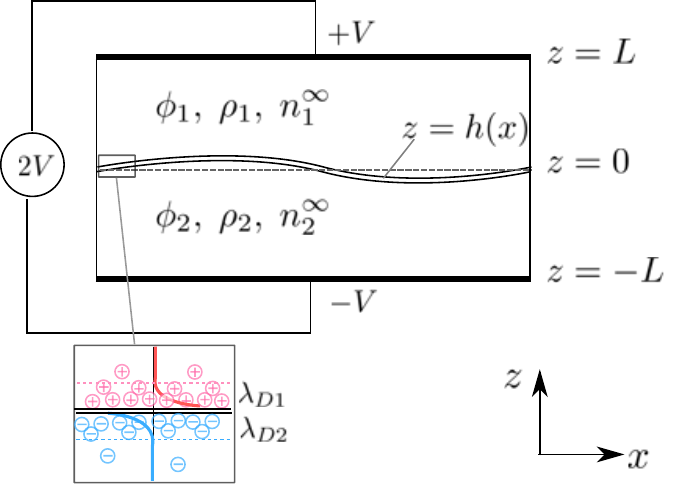}
    \caption{A planar lipid bilayer membrane, with shape denoted by $z=h(x)$, separating electrolyte solutions
with different concentrations. The electrodes, with potential $\phi=\pm V$, are held at $z=\pm L$.
    }
    \label{fig:diagram}
\end{figure}

  The system consists of two electrolyte solutions separated by a thin planar membrane impermeable to ions, as shown in Fig.~\ref{fig:diagram}. The concentrations of the positive and negative ions are $n^+_i$ and $n^-_i$, where $i=1,2$ refers to the upper fluid and the lower fluid, respectively. The electrodes are located at $z=\pm L$, where a DC voltage of $2V$ is applied. The position of the flat membrane is $z=0$. Both electrolytes are assumed to be $1:1$. The solutions have the same permittivity $\varepsilon_f$, but different concentration $n_i^\infty$. This implies different conductivities of the electrolyte solutions, $\sigma_i=2e^2\omega n_i^\infty$~\cite{Saville97_arfm}, where $\omega$ is the mobility of ions and $e$ is the elementary charge. 
  
  \zx{Throughout this study, we assume osmotic equilibration across the membrane, such that no net transmembrane water flux arises from osmotic pressure differences. In experimental systems, this condition can be achieved by adding non-permeating solutes—such as sugars like sucrose or glucose—to ensure equal osmotic pressure on both sides of the membrane~\cite{Salipante:2015}. }
 \subsection{Governing equations}

In the electrolyte, the governing equation for the electric
potential $\phi$ is the Poisson’s equation
\begin{equation}
    \nabla^2\phi_i =-\frac{\rho_i}{\varepsilon_f},
    \label{eq:possion with dimension}
\end{equation}
where $\rho_i=e(n_i^+ - n_i^-)$ is the charge density, which we assume satisfies the linearized PNP equations
 \begin{subequations}\begin{align}
&\pp{\rho_i}{t}+ \bm v_i \cdot \nabla \rho_i=-\nabla\cdot \bm j_i,\\
&\bm j_i=-D\left(\nabla\rho_i+\frac{2e^2n^\infty_i}{k_BT}\nabla\phi_i\right)\,.
\end{align}
 \end{subequations}

The perturbations in the concentration $n_i^{\pm}=n_i^{\infty}+\delta n_i^{\pm}$~\cite{Ziebert10_pre} are used to linearize the PNP, assuming small applied voltages compared to thermal voltage $k_BT/e$~\cite{Bazant04_pre}. $\bm j_i$ is the electric current density, $k_BT$ is the thermal energy, $D$ is the ion diffusion coefficient (all ions are assumed to have the same mobility), and $\bm v_i$ is the fluid velocity. Here we introduce the Debye length 
\begin{equation}
    \lambda_{D_i} =\left(\frac{2e^2n^\infty_{i}}{\varepsilon_f k_B T}\right)^{-1/2},
\end{equation}
\zxy{which} is the characteristic screening length, and we define $\lambda_{D}=(\lambda_{D_1}\lambda_{D_2})^{1/2}$ as the composite Debye length in our asymmetric system, following the notation of Ref.~\cite{Zhao25}.

In the context of biological membranes, inertia is negligible so that the velocity $\bm v_i=(u_i\hat{e}_x+v_i\hat{e}_z)$ and
pressure $p_i$ are solutions of the incompressible Stokes equations
\begin{eqnarray}
    -\nabla p_i + \mu_i\nabla^2\bm v_i -\rho_i\nabla\phi_i=0, \qquad
    \bm \nabla \cdot \bm v_i=0,
\end{eqnarray}
where $\mu_i$ is the fluid viscosity and $-\rho_i\nabla\phi_i$ is the Coulomb body force due to the electric field acting on the charge distributed in the bulk. 

 \subsection{Boundary conditions}
At the top and bottom electrodes $z=\pm L$, we impose a voltage $\phi_i(z=\pm L)=\pm V$. We neglect the polarization at the electrode surfaces~\cite{Ziebert10_pre} since we focus on the membrane dynamics. The distance between the electrodes is assumed to be much larger than the Debye length $L\gg \lambda_D$, such that the bulk electrolyte is electroneutral and thus charge density at the electrodes vanishes $\rho_i(z=\pm L)=0$.

The membrane's position is represented by $z=h(x)$, with a unit normal vector $\bm n =(-h_x,1)/(1+h_x^2)^{1/2}$ pointing upward, and a unit tangent vector $\bm \tau =(1,h_x)/(1+h_x^2)^{1/2}$. On the membrane, a Robin-type boundary condition~\cite{Lacoste09_epje,Ziebert10_pre}
 \begin{equation}
    \varepsilon_f\bm n \cdot\nabla\phi_i|_{z=h}
    =C_mV_m,\qquad i=1,2,
    \label{eq:robin with dimension}
\end{equation}
relates the transmembrane potential $V_m=\phi_1|_{z=h}-\phi_2|_{z=h}$ and the electric displacement field. $C_m=\varepsilon_m/d_0$ is the capacitance of the membrane with the thickness $d_0$ and permittivity $\varepsilon_m$.
This boundary condition was derived by mapping the continuity of the potential $\phi_i|_{z=h}=\phi_m|_{z=\pm d_0/2}$ and the continuity of the displacement field $\varepsilon_f\bm n \cdot\nabla\phi_i|_{z=h}=\varepsilon_m\bm n \cdot\nabla\phi_m|_{z=\pm d_0/2}$ to a zero-thickness membrane~\cite{Lacoste09_epje}, where $\phi_m$ is the electric field inside the membrane and $i=1 (2)$ corresponds to $+(-)d_0/2$. The continuity of the displacement field holds under the assumption of no adsorbed charges at the membrane interface. It is important to note that while the charge in the Debye layer at the membrane may be nonzero, the surface charge is zero.
In fact, the bulk charge integrated over the Debye layer in the electrokinetic model~\cite{Zhao25}, expressed as $Q=\int_{\lambda_D}\rho dz$, corresponds to the surface charge in the leaky dielectric model. For a flat membrane, we calculate the macroscopic charge from Eq.~\eqref{eq:possion with dimension}:
\begin{multline}
    Q_1=\int_{z=h}^{z=+\infty}\rho_1 dz\\
    =\left.\varepsilon_f\pp{\phi_1}{z}\right|_{z=h}
    -\left.\varepsilon_f\pp{\phi_1}{z}\right|_{z=\infty}
    \approx \left.\varepsilon_f\pp{\phi_1}{z}\right|_{z=h},
    \label{eq:macroscopic charge}
\end{multline}
since $z=+\infty$ is taken as the position far from the membrane, and electric field vanishes in the bulk $-\pp{\phi_1}{z}|_{z=+\infty}=0$. 
The macroscopic charge on the other side of the membrane, denoted as $Q_2$, is calculated similarly. Upon analysis, the Robin \zxy{condition} in Eq.~\eqref{eq:robin with dimension} further reveals $Q_1=-Q_2$, signifying that the total charge on both sides of the membrane always remains balanced, regardless of conductivity asymmetry. 

Another boundary condition is the no-flux condition~\cite{Mori18_jfm}:
\begin{equation}
    \bm n \cdot \bm j_1|_{z=h}= \bm n \cdot\bm j_2|_{z=h}=0.
    \label{eq:non-flux}
\end{equation}
This corresponds to an insulating membrane, i.e., charge is not permitted to cross the membrane. 

For the flow, the no-slip and no-penetration boundary condition applies for the velocity field at the electrodes $\bm v (z=\pm L)=\bm 0$. The area-incompressibility of the membrane implies that the surface velocity is solenoidal, $\nabla_s\cdot \bm v_s=0$, where $\nabla_s=(\bm I-\bm n \bm n )\cdot \nabla$ is the surface gradient operator. The normal stress balance on the interface is given by:
 \begin{eqnarray}
\ljump
\bm n\cdot \bm{T}\cdot \bm n\rjump_h&=&\bm n  \cdot \bm \tau_m,
 \end{eqnarray}
 where we denote $\ljump f\rjump_a=f_1(z=a)-f_2(z=a)$ as the jump of $f$ at position $z=a$. $\bm T$ is the total stress, consisting of the hydrodynamic stress ${\bm T^{hd}}$ and Maxwell's stress ${\bm T^{el}}$ 
 \begin{equation}
\bm{T}
=\underbrace{-p \bm{I} +
\mu[\bbm\nabla \bv+(\bbm\nabla \bv)^\top]}_{\bm T^{hd}}
\underbrace{+\varepsilon_f\left[\bm{E} \bm{E}\\
-(\bm{E}\cdot\bm{E})\bm{I}/2\right] }_{\bm T^{el}}.
\end{equation}
The stresses jump across the membrane is balanced by membrane elastic stresses derived
from the Helfrich energy~\cite{Seifert99_EPJ}:
\begin{multline}
    \bm \tau_m=-\kappa[(2H-C_0)(2H^2-2K_G+C_0H)+2\nabla_s^2H]\bm n\\
    +2\Gamma H\bm n-\nabla_s\Gamma,
\end{multline}
where $H$ is the curvature, $K_G$ is the Gaussian curvature, and $\Gamma$ is the membrane tension. Note that for \zxy{an} incompressible membrane, the tension is determined by the membrane’s mechanical response~\cite{Young14_fjm}. The tangential electric stresses are balanced by gradients in the membrane tension, such that non-uniform tension arises to enforce local area incompressibility. In the linearized formulation, however, the tension remains uniform at leading order, and variations appear only at higher order~\cite{Young14_fjm}. $C_0$ is the bilayer spontaneous curvature arising from the asymmetry of the monolayers. In this stability study, we initially neglect $C_0$ by assuming $C_0d_0\ll1$~\cite{Seifert93_el} for a membrane of zero thickness. We come back to this effect in Sec.~\ref{sec:curvature}. 

Finally the kinematic condition determines the evolution of the interface:
\begin{eqnarray}
\zx{ \pp{h}{t}=v-u\pp{h}{x}.}
\end{eqnarray}
 \subsection{Dimensionless parameters and rescaling}
All variables are scaled using their characteristic values:
 \begin{eqnarray}
\rho_i&\mapsto&  2e\sqrt{n_1^\infty n_2^\infty}\rho_i,\quad  \phi_i\mapsto \underbrace{\frac{k_BT}{e}}_{\phi_c}\phi_i,\quad  V\mapsto \phi_c V,\nonumber\\
(x,z)&\mapsto&  \lambda_D(x,z),\quad  h\mapsto \lambda_D h,\quad d_0\mapsto\lambda_D d,\\
t&\mapsto& \underbrace{\frac{\mu \lambda_D^2}{\varepsilon_f\phi_c^2}}_{t_{el}}t
,\quad
p\mapsto\frac{\mu}{t_{el}} p,\nonumber
 \end{eqnarray}
where $\phi_c$ is the thermal voltage, and $t_{el}$ is the flow time scale~\cite{Seiwert12_jfm}. Using this time scale, the scaled PNP model takes the form:
 \begin{subequations}\label{eq:pnp possion dmsless} \begin{align}
     \alpha\left(\pp{\rho_1}{t}
          +\bm v_1\cdot \nabla \rho_1\right)
     &=\nabla^2 \rho_1 
     +R^{1/2}\nabla^2\phi_1,\label{eq:PNP dmsless} \\
     \nabla^2\phi_{1}&=-\rho_{1},\label{eq:possion dmsless}
\end{align}
\end{subequations}
where 
 $R=\frac{\sigma_1}{\sigma_2}=\frac{n_1^\infty}{n_2^\infty}$ denotes the conductivity ratio. With $R\mapsto R^{-1}$, Eqs.~\eqref{eq:pnp possion dmsless} applies for phase 2.
 The parameter $\alpha$ represents the ratio between $t_{el}$ and the characteristic charging time of the double-layer capacitor $t_c=\frac{\lambda_DL}{D}$~\cite{Bazant04_pre,Zhao25}:
 \begin{equation}
     \alpha=\frac{\lambda_D^2}{D}\frac{1}{t_ {el }}=
     \frac{1}{D}\frac{\varepsilon_f\phi_c^2}{\mu }=
     \frac{t_c}{ t_{el}}\frac{\lambda_D}{L}.
 \end{equation}
 In this study, we consider the situation where $\alpha\ll1$, and set $\alpha=0$. In this limit, the charge relaxation in Debye layer happens instantaneously, and the Debye layer remains at equilibrium while the membrane is deforming.
The dimensionless Robin \zxy{condition} is given by
 \begin{equation}
    \bm n \cdot\nabla\phi_i|_{z=h}
    =\beta V_m|_{z=h},\qquad i=1,2,
    \label{eq:robin without dimension}
\end{equation}
where 
\begin{equation}
   \beta=\frac{C_m \lambda_D}{\varepsilon_f}=\frac{\varepsilon_m}{\varepsilon_fd} 
   \label{eq:beta}
\end{equation}
is the dimensionless capacitance. The no-flux condition at the membrane retains the form in Eq.~\eqref{eq:non-flux}
 with the dimensionless current
  \begin{equation}
\bm j_1=-\nabla(\rho_1 +\gamma^{-2}\phi_1),\qquad
\bm j_2=-\nabla( \rho_2 +\gamma^{2}\phi_2),
 \label{eq:current without dimension}
 \end{equation}
 where we denote $\gamma=R^{-1/4}$ for convenience.
The scaled Stokes equation is given by
   \begin{eqnarray}
    -\nabla p + \nabla^2\bm v -\rho\nabla\phi=0.
    \label{eq:scaled stokes}
\end{eqnarray}
 The scaled normal stress balance on the interface is
\zx{  \begin{multline}
        \ljump -p+2 \partial_nv_n \\
        +\frac{1}{2}[(\partial_n\phi)^2-(\partial_\tau\phi)^2]
         -\frac{1}{2}\beta d[(\partial_n\phi_m)^2-(\partial_\tau\phi_m)^2]
       \rjump _{z=h}\\
    =-\frac{1}{Ca}(4H^3-4HK_G+2\nabla_s^2H-2\bar\Gamma H),
\label{eq:scaled normal stress}
  \end{multline}   
where $\partial_nv_n=\bm n\cdot [\bbm\nabla \bv+(\bbm\nabla \bv)^\top]\cdot \bm n$ is the projection of the viscous stresses, and $\partial_\tau\phi=\bbm\nabla\phi \cdot \bm \tau$.} The capillary number $Ca$ compares the electric and bending stresses: 
    \begin{eqnarray}
    Ca=\frac{\lambda_D\varepsilon_f\phi_c^2}{\kappa},\quad \bar\Gamma=\frac{\Gamma \lambda_D^2}{\kappa}.
    \end{eqnarray}
\zx{In Sec.~\ref{sec:solution}, we will show how the electric field impacts the force balance on the membrane via the Maxwell stress and the \zxy{hydrodynamic} pressure $p$.}
The traction from the Maxwell stress at the membrane consists of two parts, with $\ljump\frac{1}{2}(\partial_z\phi)^2\rjump$ coming from the electric field in the external medium, and $\ljump -\frac{1}{2}\beta d(\partial_z\phi_m)^2\rjump$ coming from the electric field $\bm E_m=-\nabla \phi_m$ in the membrane~\cite{Isambert98_prl,lacoste07_el}. We emphasize that although the Robin condition in Eq.~\eqref{eq:robin without dimension} effectively maps the electric field and charge density on both sides of a finite-thickness membrane to a zero-thickness membrane, the $\phi_m$ term in Eq.~\eqref{eq:scaled normal stress} remains necessary to calculate the effective stress in the zero-thickness limit~\cite{Lomholt07_pre,Ziebert10_pre,Loubet13_pre}. We will address this in Sec.~\ref{sec:solution}.

 \section{Linear stability analysis}
  \label{sec:solution}
The membrane fluctuations impose small perturbation $h(x,t)= h_k e^{st+ikx}$ around a flat reference state. Any variable $g$ is expanded in a series of the form $g=g^{(0)}(z,t)+g^{(1)}(z,x,t)$. The superscript $(0)$ corresponds to the base state, and $g^{(1)}=  g_k^{(1)} e^{st+ikx}$ is the first order solution due to the effect of the small undulations. For convenience, we drop the subscript $k$ for the first order perturbation.
\subsection{Base state}
We consider the flat membrane as the base state, $h(x)=0$, and the
electric field is aligned in $z$ direction, perpendicular to the membrane. For electrostatic fields, we solve Eq.~\eqref{eq:PNP dmsless} and Eq.~\eqref{eq:possion dmsless}, with the Robin \zxy{condition}~\eqref{eq:robin without dimension} and no-flux condition~\eqref{eq:non-flux} at the membrane, and $\rho_i(z\rightarrow \pm \infty)=0$, $\phi_i(z\rightarrow \pm \infty)=\pm V$ at the far field, to obtain the charge distribution
  \begin{eqnarray}
\rho_1^{(0)}=\rho_{m}e^{-\frac{z}{\gamma}},\qquad
\rho_2^{(0)}=-\gamma^2\rho_{m}e^{\gamma z},
 \end{eqnarray}
 and the base state electric potential
  \begin{eqnarray}
\phi_1^{(0)}=-\gamma^2\rho_{m}e^{-\frac{z}{\gamma}}+V,\qquad
\phi_2^{(0)}=\rho_{m}e^{\gamma z}-V,
 \end{eqnarray}
 where 
  \begin{eqnarray}
     \rho_{m}=\frac{2\beta V}{\gamma+(1+\gamma^2)\beta}
 \end{eqnarray}
is related to the jump in the charge density across the membrane, with $\ljump\rho^{(0)}\rjump_{z=0}=(1+\gamma^2) \rho_{m}$. Note that in the base state, the current $\bm j_1^{(0)}=\bm j_2^{(0)}=\bm 0$ for all $z$, not just at the membrane~\cite{Zhao25}.

The base state electric field calculated as $\bm E_i^{(0)} =E_{z,i}^{(0)} \bm{e}_z=
-\partial_z \phi_i^{(0)} \bm{e}_z$, is non-zero in the Debye layer. This represents a key difference from the base state field in LDM. In LDM, under the same physical setup, the base state field is zero unless there is ion current across the membrane~\cite{Seiwert12_jfm,Sens02_prl} or membrane capacitor is charging~\cite{Schwalbe11_pof}. In our subsequent discussion, we will demonstrate the critical role played by the non-zero base state electric field $\bm E_i^{(0)}$ in the membrane instability.
The transmembrane potential in the base state is given by
\begin{equation}
V_m^{(0)} =\ljump\phi^{(0)}\rjump_{z=0}=\frac{2V\gamma}{\gamma+(1+\gamma^2)\beta}.
\label{eq:transmembrane potential}
\end{equation}
Correspondingly, the internal field $E_m^{(0)}$ is
\begin{equation}
    E_m^{(0)}=-V_m^{(0)}/d=-\frac{\rho_m}{\beta}\frac{\gamma}{d},
\end{equation}
where the constant internal field has been used due to the continuity of the potential at the membrane boundaries, which is consistent with $\varepsilon_m E_m^{(0)}=\varepsilon_fE_{z,i}^{(0)}|_{z=0}$ by substituting $\beta$ in Eq.~\eqref{eq:beta}. The internal field $E_m^{(0)}$ is obtained by keeping a finite thickness $d$, and in Sec.~\ref{sec: linear growth rate}
we will show this field introduces negative tension and results in a destabilization effect. 

 In the bulk fluid, the velocity is zero $\bm v_i^{(0)}=\bm 0$, and the base-state pressure is obtained from the Stokes equation~\eqref{eq:scaled stokes} as 
\begin{equation}
     p^{(0)}_i =\frac{1}{2}\left(\pp{\phi_i^{(0)}}{z}\right)^2
     =\frac{1}{2}\gamma^2\rho_m^2e^{\mp 2z\gamma^{\mp 1}}
    \label{eq:pressure p0}
\end{equation}
by assuming $p_i^{(0)}(z\rightarrow \pm \infty)=0$. This indicates that the pressure is continuous $\ljump p^{(0)}\rjump_{z=0}=0$ across the flat membrane. \zx{Equation~\eqref{eq:pressure p0} also implies that in the bulk fluid there is a balance between the pressure and the Maxwell stress in the absence of viscous stress, i.e., $\nabla\cdot \bm T=0$.
}
  \subsection{Perturbed state equations}
In the perturbed state, by setting $\alpha=0$,  the ion conservation and Poisson’s equation become
\begin{subequations}\begin{align}
(\partial_z^2 -{k}^2 )\left(\rho_{1}^{(1)}+\gamma^{-2}\phi_{1}^{(1)}\right)&=0, \\  
       (\partial_z^2 -{k}^2 )\left(\rho_{2}^{(1)}+\gamma^{2}\phi_{2}^{(1)}\right)&=0,\\
      \label{eq:leading ion conservation}
(\partial_z^2 -{k}^2  )\phi_i^{(1)}+\rho_i^{(1)}&=0,
\end{align}
\label{eq:leading Poisson}
\end{subequations}
which is solved to yield the first order potential 
 \begin{equation}
     \phi_1^{(1)}=-\rho_{m}a_1he^{-l_{1}z},\qquad
     \phi_2^{(1)}=-\rho_{m}a_2he^{l_{2}z},
      \label{eq:phi k}
 \end{equation}
 and charge density 
 \begin{equation}
       \label{eq:rho k}
     \rho_{1}^{(1)}=-\gamma^{-2}\phi_{1}^{(1)}, \quad\rho_{2}^{(1)}=-\gamma^{2}\phi_{2}^{(1)}.
 \end{equation}
 The $l_i$ represent the inverse characteristic length scale of the electrostatic potential near the slightly undulated membrane, with
 \begin{eqnarray}
      l_1=(\gamma^{-2}+{k}^2)^{1/2},\quad l_2=(\gamma^{2}+{k}^2)^{1/2}.
 \end{eqnarray}
 The coefficients 
    \begin{equation}                                                    a_1=
     \frac{(\gamma^2+1)\beta + l_2}{l_1l_2 +\beta(l_1+l_2)}
     ,\quad
     a_2=
     \frac{(\gamma^2+1)\beta +\gamma^2 l_1}{l_1l_2 +\beta(l_1+l_2)}
     .
     \label{eq:Ar Al}
 \end{equation}
are obtained via imposing the first order of Robin \zxy{condition} at the membrane
 \begin{eqnarray}
     \left( h\partial_{zz}\phi_{i}^{(0)}+\partial_{z}\phi_{i}^{(1)}  \right)_{z=0} 
     =\beta\ljump \phi^{(1)} + h\partial_{z}\phi^{(0)}\rjump _{z=0}.
     \label{eq:Robin BC first}
 \end{eqnarray}
Note that, when subjected to perturbation, the first order current remains zero in both electrolytes $\bm j_1^{(1)}=\bm j_2^{(1)}=\bm 0$.
 
To obtain the flow induced in the surrounding electrolyte due to the perturbation, we solve the first order incompressibility and the
Stokes equation 
\begin{subequations}\begin{align}
    \partial_zv^{(1)}_i+iku^{(1)}_i&=0,\label{eq:continiuty k}\\
    -\partial_z p^{(1)}_i+(\partial_z^2-{k}^2)v^{(1)}_i -(\rho^{(1)}_i\partial_z\phi^{(0)}_i+\rho^{(0)}_i\partial_z\phi^{(1)}_i)&=0,\label{eq:LM-x k}\\
     -ik p^{(1)}_i+(\partial_z^2-{k}^2)u^{(1)}_i-(ik\rho^{(0)}_i\phi^{(1)}_i)&=0.\label{eq:LM-z k}
\end{align}
\label{eq:continiuty LM k}
\end{subequations}
Rearranging Eqs.~\eqref{eq:continiuty LM k} gives rise to the following fourth-order differential equation for $v_i$:
 \begin{equation}
     (\partial_z^2-{k}^2)^2v^{(1)}={k}^2(\phi^{(1)}\partial_z\rho^{(0)}-\rho^{(1)}\partial_z\phi^{(0)})=0.\label{eq:vel fourth}
 \end{equation}
  
 By imposing $v_i(z\rightarrow\pm \infty)=0$, we obtain the solution of the form $v_i^{(1)}=(B_i+C_iz)e^{\mp kz}$, where $B_i$ and $C_i$ are to be determined by appropriate boundary conditions. The continuity of the tangential velocity, together with the incompressibility of the interface at the first order impose $u_1^{(1)}|_{z=0}=u_2^{(1)}|_{z=0}=0$. This reduces the kinetic equation to be $v^{(1)}=\partial_t h=sh$, together with the continuity of the normal velocity we have $v_1^{(1)}|_{z=0}=v_2^{(1)}|_{z=0}=sh$. These boundary conditions determine the solutions as function of $s,h$. \zx{Note that, unlike the induced-charge electro-osmotic (ICEO) mechanism identified in Refs.~\cite{Ziebert10_pre,Lacoste09_epje}, the flow field in the present study is not directly driven by the electric field. Instead, it is the deformation and motion of the membrane that generates the surrounding fluid flow.
 }

The pressure $p^{(1)}_i=(\partial_z^3/{k}^2-\partial_z)v_i^{(1)} -(\rho^{(0)}\phi^{(1)})$ is obtained as 
   \begin{equation}
       p_1^{(1)}=2kshe^{-kz}-\rho_{1}^{(0)}\phi_{1}^{(1)},\qquad
       p_2^{(1)}=-2 kshe^{kz}-\rho_{2}^{(0)}\phi_{2}^{(1)},
       \label{eq:pressure k}
   \end{equation}
\zx{which includes both the dynamic pressure associated with the fluid flow and electrostatic pressure due to the body force.}
To determine the growth rate $s$, we apply the normal traction balance at the interface
    \begin{multline}   
        \bigg[\!\!\bigg[ h\partial_z\left[-p^{(0)}+\frac{1}{2}(\partial_z\phi^{(0)})^2\right]
        -p^{(1)}+2 \partial_zv^{(1)} +\partial_z\phi^{(0)}\partial_z\phi^{(1)}\\
        -\beta d\partial_z\phi_m^{(0)}\partial_z\phi_m^{(1)}
       \bigg]\!\!\bigg] _{z=0}
        =\frac{2}{Ca}(\bar\Gamma{k}^2+{k}^4)h.
        \label{eq:normal traction k}
    \end{multline}
The term $ h\partial_z\left[-p^{(0)}+\frac{1}{2}(\partial_z\phi^{(0)})^2\right]$ vanishes due to the stress balance at the base state, and $  \partial_zv^{(1)}=0$ due to the incompressibility of the membrane and the bulk fluids. From Eq.~\eqref{eq:normal traction k} we see three sources of electrostatic contribution to the membrane instability, i.e., one from the pressure difference $\ljump -p\rjump$ due to the \zx{electrical body force}, one from the external normal traction $\ljump \partial_z\phi^{(0)}\partial_z\phi^{(1)}\rjump$, and one from the internal normal traction $\ljump -\partial_z\phi_m^{(0)}\partial_z\phi_m^{(1)}\rjump$ of the membrane. Note that while the base state field $-\partial_z\phi_m^{(0)}$ is constant, the first order field $-\partial_z\phi_m^{(1)}$ varies across the membrane (see Appendix~\ref{app:internal normal traction} and Ref.~\cite{Ziebert10_pre} for detailed calculation). For a zero-thickness model, this internal traction is needed to correct the electrohydrodynamic effect as seen in Ref.~\cite{Ziebert10_pre} 
and Ref.~\cite{Loubet13_pre}.
\subsection{Linear growth rate}
\label{sec: linear growth rate}
\begin{figure}[ht!]
    \centering
\includegraphics[keepaspectratio=true,width=\columnwidth]{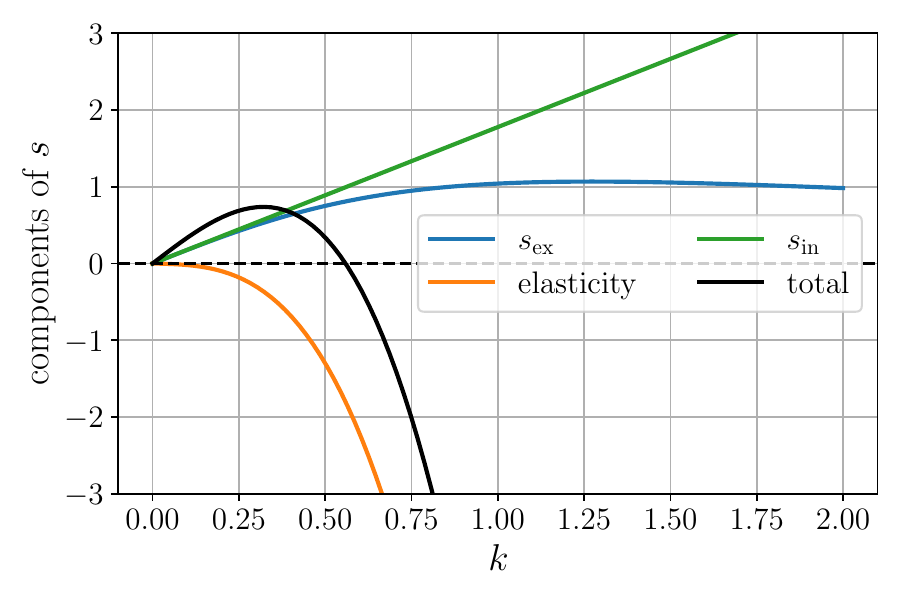}
    \caption{
    The components of the growth rate in Eq.~\eqref{eq:growth rate full} as a function of the perturbation wavenumber $k$, with the ``total" representing the sum of the external, internal, and elastic contributions.
    We consider symmetric conductivity with $\gamma=1$, $V=2$, \zx{$\beta=1$, $Ca=0.2$, $\bar \Gamma=0.01$} and $d=0$.
    }
    \label{fig:s component}
\end{figure}

The growth rate is determined from Eq.~\eqref{eq:normal traction k} as
 \begin{equation}
    4s= s_{\text{ex}}+ s_{\text{in}}
    -\frac{2}{Ca}\left(\bar\Gamma k+ {k}^3\right),
    \label{eq:growth rate full}
 \end{equation}
where $ s_{\text{ex}}$ represents the external contribution to the growth rate, including both \zx{electrical body force} and normal traction \zx{from the Maxwell stress}, whereas the internal contribution $ s_{\text{in}}$ arises solely from normal traction:
 \begin{subequations}\label{eq:ex in}\begin{align}
 s_{\text{ex}}&=\frac{\rho_{m}^2}{ k}[
    \underbrace{-(\gamma^2a_{2}+a_{1})}_{\text{electrical body force}}+
    \underbrace{\gamma(\gamma^2+1)}_{\text{\zx{Maxwell stress traction}}}
    ],\label{eq:external}\\
 s_{\text{in}}&=\frac{\rho_{m}^2\gamma^2}{\beta }\left[2-\frac{\beta d}{\gamma}(2\gamma-a_1-a_2)\right]\frac{e^{kd}-1}{(e^{kd}+1)d}.\label{eq:internal}
 \end{align}\end{subequations}
 The three components, along with the full growth rate in Eq.~\eqref{eq:growth rate full} are shown in Fig.~\ref{fig:s component}. Note that in the zero thickness limit $d\rightarrow 0$, the internal normal traction term in Eq.~\eqref{eq:internal} reduces to 
\begin{equation}
    \lim_{d\rightarrow 0}s_{\text{in}}
    =\frac{\rho_{m}^2\gamma^2}{\beta }k,
\end{equation}
 contributing linearly to the growth rate. This term is in agreement with Ref.~\cite{Sens02_prl} in the LDM framework.
  
Next, we examine the effect of each term in the long-wave limit $k\rightarrow 0$. With a finite thickness $d$, the internal normal traction is expanded as 
\begin{equation}
        \lim_{k\rightarrow 0}s_{\text{in}}
    =\frac{\rho_{m}^2}{\beta }
    \left[\gamma^2k
    -\left(\frac{d}{2}\beta \gamma m_2+\frac{d^2}{12}\gamma^2\right)k^3
    \right]+O(k^5),
\end{equation}
with  $ m_2=\gamma+\left(\gamma^2 - 1\right)^{2} /[2 \left(\beta \gamma^{2} + \beta + \gamma\right)]$.

The \zx{electrical body force} in Eq.~\eqref{eq:external} is \zxy{expanded} as
\begin{equation}
-(\gamma^2a_{2}+a_{1})
=-\gamma(\gamma^2+1)
+s_2 {k}^2
-s_4{k}^4
+O({k}^6),
\label{eq:body force expansion}
\end{equation}
with 
 \begin{subequations} \begin{align}
    s_{2}
     &=\frac{\gamma(\gamma^2+1)}{2},\nonumber\\
     s_{4}&=
     \frac{ \gamma^{2} + 1}{8\gamma}\left(\gamma^{4} + \gamma^{2} + 1+ \frac{2 \gamma \left(\gamma^{2} - 1\right)^{2}}{\beta \gamma^{2} + \beta + \gamma} \right).\nonumber
 \end{align}\end{subequations} 
Note that the constant term $-\gamma(\gamma^2+1)$ in Eq.~\eqref{eq:body force expansion} is canceled by the external normal traction \zx{from the Maxwell stress} in Eq.~\eqref{eq:external}, and the higher-order terms of $k$ \zx{represent the amount that the external normal traction from the Maxwell stress exceeds the electrical body force.}

Thus, in the long-wave limit the growth rate has the form
    \begin{multline}       
    4s= -\left(\frac{2}{Ca}\bar\Gamma-\rho_{m}^2s_2
   - \rho_m^2\frac{\gamma^2}{\beta}
    \right)k\\
     -\left[\frac{2}{Ca}+\rho_{m}^2 s_4
     +\frac{\rho_m^2\gamma^2}{\beta}\left(
  \frac{d^2}{12}+\frac{m_2}{2\gamma}
  \beta d
  \right)
     \right]{k}^3.
         \label{eq:growth rate long}
 \end{multline}
\zx{The electrostatic contributions from the external term, $\rho_m^2 s_2 > 0$, and the internal term, $\rho_m^2 \frac{\gamma^2}{\beta} > 0$, act to destabilize the system by effectively reducing the membrane tension. This reduction is interpreted as a negative correction to the membrane tension—an effect analogous to the classical Lippmann (electrocapillary) phenomenon~\cite{zhang01_nature}.} For the cubic contribution, the external term $\rho_{m}^2 s_4>0$ and the internal term $\frac{\rho_m^2\gamma^2}{\beta}\left(\frac{d^2}{12}+\frac{m_2}{2\gamma}\beta d \right)>0$ enhance the stabilizing effect of the bending modulus at shorter wavelength (greater $k$).
When conductivity is symmetric on both sides, i.e. $\gamma=1$, we reproduce the growth rate in Ref.~\cite{Ziebert10_pre}, and the electrostatic corrections to the tension and bending modulus in Ref.~\cite{Loubet13_pre,lacoste07_el}.
Such linear destabilizing effect is also reported in Ref.~\cite{kumaran2000_prl} as a consequence of the surface charge discrepancy between the two lipid layers for a \zxy{tension-free} membrane.

\zx{
Note that the electrostatic correction to the tension, represented by the term $\rho_m^2 s_2$, arises from a partial cancellation between the stabilizing electrical body force and the destabilizing Maxwell stress at the membrane, as seen from Eq.~\eqref{eq:body force expansion}. Specifically, the negative contribution from the body force and the positive traction due to the Maxwell stress together determine the net effect on the growth rate~\eqref{eq:external}. Since the Maxwell stress dominates in this balance, the remaining contribution $\rho_m^2 s_2$ acts to destabilize the membrane. The physical mechanisms underlying these opposing effects are discussed in detail in Sec.~\ref{sec:effect of forces}.
}

The instability observed in the present system, as reported by this study and Ref.~\cite{Ziebert10_pre}, appears to be independent of the ion current or transient effects. This presents a significant departure from the behavior predicted by the LDM model in Ref.~\cite{Seiwert12_jfm}, where, under identical conditions, the fully charged membrane remains stable. This \zxy{difference} in stability outcomes—despite originating from the same physical setting—raises an important question about the mechanisms responsible for the onset of instability. We address these differences and clarify the underlying factors governing membrane stability in the next section.

\section{Instantaneous charge relaxation with asymmetric membrane}
\label{sec: alpha 0}
In this section, we first examine the effects of key parameters, followed by a discussion of the stabilizing electrostatic body forces and destabilizing electrostatic surface traction separately. Next, we show that asymmetric electrolytes conductivities give rise to a membrane spontaneous curvature. Finally, we compare the linear stability analyses based on LDM and EK frameworks.

We consider the physical \zx{parameters relevant to} cells or giant lipid vesicles, and thus the fluids are aqueous salt solutions or water. Our analysis focuses on the case of equal-density and equal-viscosity fluids, $\rho_1=\rho_2$ and $\mu_1=\mu_2\approx 10^{-3}$ Pa s. The value of the Debye length varies from $\lambda_D\sim 1 \;\mathrm{\mu m}$ for pure water to \zx{$\lambda_D\sim 1 \;\mathrm{nm}$ for 100 mM NaCl.} \zx{This range of ionic strengths reflects physiological conditions, is commonly used in biomimetic membrane experiments~\cite{riske2005electro,Salipante:2012,salipante2014vesicle,Salipante:2015}, and enables the investigation of asymmetric ionic conductivity observed in experimental systems~\cite{aranda2008morphological}.}
For pure water, the absolute permittivity is around $\varepsilon_f\approx7\times10^{-10}\;\mathrm{F\:m^{-1}}$. The diffusion coefficient is around $D\sim 10^{-9}\; \mathrm{m^2\:s^{-1}}$.
The character length scale for the gap between the electrodes is $L\sim 100\; \mathrm{\mu m}$. For the lipid membrane, typical values for the thickness are around $d_0\sim5\;\mathrm{nm}$, and the capacitance is around $C_m\sim 0.01 \;\mathrm{F\; m^{-2}}$. The elastic behavior is generally characterized by the bending modulus around $\kappa\sim 10^{-20}\;\mathrm{J}$. The typical value for the stretching modulus of a lipid membrane is around \zx{$\Gamma\sim 10^{-5}\; \mathrm{J\;m^{-2}}$~\cite{Lacoste09_epje}}, and the electric field has additional effects on it, such as the compressive electric stress effectively decreasing the isotropic tension~\cite{Needham89_bj,Ziebert10_pre}. 
From these typical values we obtain other dimensionless numbers as \zx{$Ca\sim 0.01-10$, $\bar \Gamma=10^{-3}-10^3$ and $\beta\sim 0.01-10$}.
The time scale ratio is around $\alpha\sim 0.1$, and we consider the instantaneous charge relaxation, i.e. $\alpha=0$ in this work. The conductivity ratio $R$ can vary over a broad range~\cite{Seiwert12_jfm}.

\subsection{\zxy{Effect of physical parameters}}
\begin{figure}[ht!]
    \centering
\includegraphics[keepaspectratio=true,width=0.9\columnwidth]{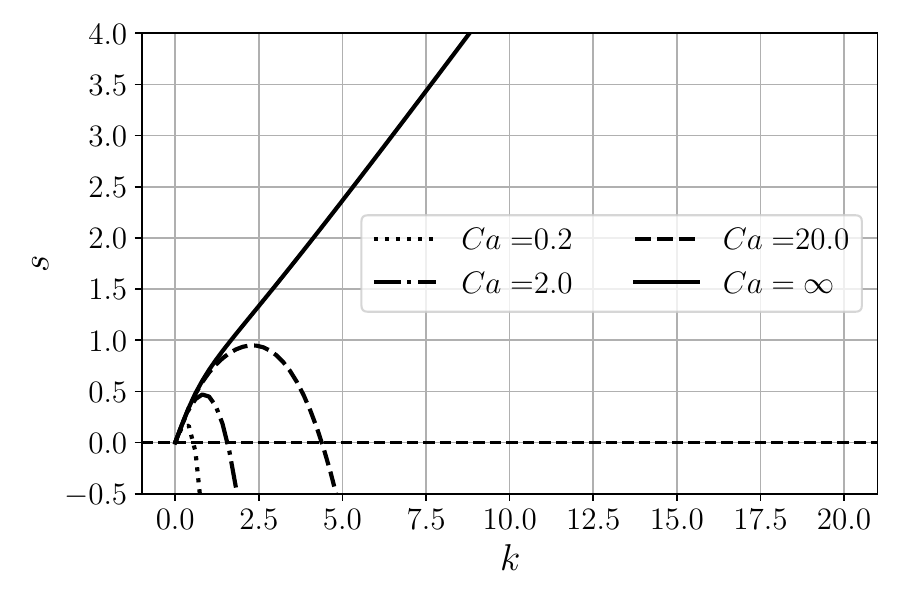}\put(-30,130){(a)}

\includegraphics[keepaspectratio=true,width=0.9\columnwidth]{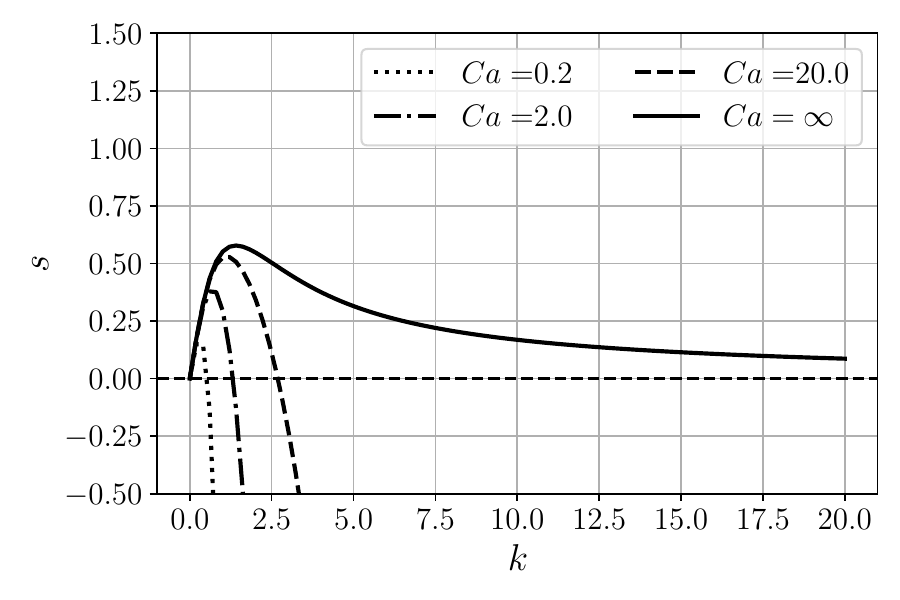}\put(-30,30){(b)}
    \caption{
    The effect of $Ca$ on growth rate $s$ as a function of perturbation wavenumber $k$. We consider symmetric conductivity with $\gamma=1$, $V=2$, \zx{$\beta=1$, $\bar\Gamma=0.01$} for (a) $d=0$ and (b) $d=1$.
    }
    \label{fig:ca effect}
\end{figure}

The growth rate in the long-wave limit Eq.~\eqref{eq:growth rate long}, i.e. $k\rightarrow 0$, shows the electrostatic forces serve as the correction to the membrane tension and bending modulus, with the tension destabilizes the system at the leading order, i.e., linear in $k$. The perturbation with smaller wavelength is dominated by the stabilizing effect of the bending elasticity. Thus, the system is unstable to a finite window of wavenumbers $0<k<k_c$, where $k_c$ is the critical wavenumber corresponding to the marginal stable state $s=0$ as the root of Eq.~\eqref{eq:growth rate full}.

The effect of $Ca$ is illustrated in Fig.~\ref{fig:ca effect}. The growth rate monotonically increases with $Ca$. As $Ca\rightarrow +\infty$, the elastic effect vanishes and all wavelengths are unstable due to electrostatic forces. In the zero-thickness limit, the growth rate at larger $k$ is dominated by the linear behavior of the internal normal traction for $Ca\rightarrow +\infty$, while for a finite thickness membrane, the growth rate first increases then saturates to zero for $Ca\rightarrow +\infty$, which is similar to the description in LDM~\cite{Seiwert12_jfm}. However, it is noteworthy that the length scale in this work is the Debye length, and thus $k\rightarrow\infty$ represents the perturbation with wavelength much smaller than Debye length, which differs from Ref.~\cite{Seiwert12_jfm}, where short-wave limit $k\rightarrow\infty$ is scaled with the inverse geometric length. 

\begin{figure}[ht!]
    \centering
\includegraphics[keepaspectratio=true,width=0.95\columnwidth]{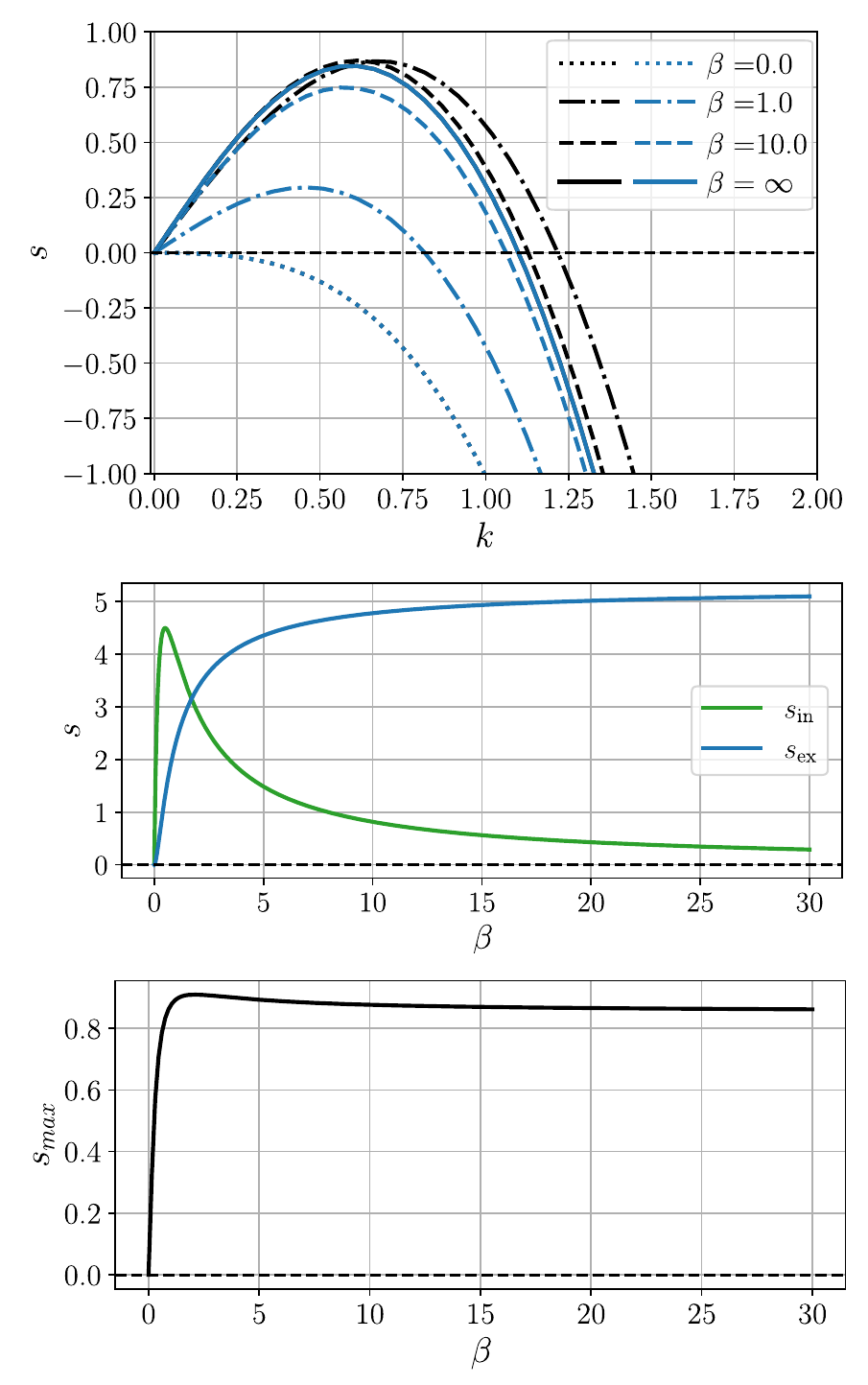}\put(-30,250){(a)}\put(-30,195){(b)}\put(-30,35){(c)}
    \caption{
(a)The effect of $\beta$ on the growth rate, with black curve shows the full growth rate as in Eq.~\eqref{eq:growth rate full}, while the blue curve excludes the contribution from the internal traction in Eq.~\eqref{eq:internal}; (b) the effect of $\beta$ on the external and internal contribution to the growth rate in Eq.~\eqref{eq:ex in} at $k=1$;
(c) the maximal (full) growth rate $s_{\max}$ as a function of $\beta$. Note that the blue and black curves overlap at $\beta=0,\infty$ in (a). We consider the system with the parameters \zx{$V=3$, $\gamma=1$, $d=0$, $Ca=0.5$, and $\bar \Gamma =0.01$.}
    }
    \label{fig:beta effect}
\end{figure}

It is interesting to note that the effect of $\beta$ in Fig.~\ref{fig:beta effect}(a) on the growth rate is very similar to that of the conductivity of the membrane, $g_m$ in Ref.~\cite{Seiwert12_jfm}, Fig.~2(a), where $g_m=G_mL/\sigma$ and $G_m$ is the dimensional membrane conductivity. At $\beta \;(g_m)=0$, the system is stable for all wavenumbers, while at $\beta \;(g_m)\rightarrow+ \infty$ the growth rate saturates. Mathematically, such similar behavior can be understood from the continuous current condition at the membrane for steady state in Ref.~\cite{Seiwert12_jfm} (see Eq.~(2.2)):
\begin{equation}
    \bm n \bm\cdot \bm E_i=g_m V_m,\qquad i=1,2,
    \label{eq:current}
\end{equation}
which is in the same form as the Robin condition~\eqref{eq:robin without dimension} at the membrane: $\bm n \bm\cdot  (-\bm E_i)=\beta V_m$, where $\beta=C_m \lambda_D/ \varepsilon_f$. The difference in the sign of the electric field, i.e. $-\bm E_i$ is due to the different relative direction of the surface normal $\bm n$ w.r.t. the potential at electrodes. Moreover, in these two limits, the internal normal stress is zero, and the system in the current work has the same stress boundary at the membrane as in Ref.~\cite{Seiwert12_jfm}.
Thus, even if the dimensionless number $g_m$ and $\beta$ \zxy{have} very different physical meanings, and the membrane boundary condition Eq.~\eqref{eq:robin without dimension} and Eq.~\eqref{eq:current} represent balance of the charge near membrane and continuity of the current respectively, they do have the similar effect on the stability of the fully charged membrane mathematically. 

Next, we illustrate the physical meaning at two limits $\beta\rightarrow 0, +\infty$. As $\beta\rightarrow 0$, there is no charge accumulated in the Debye layer, i.e., the Debye layer vanishes. The base state potential are constant in space as $\phi^{(0)}_i(z)=\pm V$, and the electric field $E_{i,z}^{(0)}(z)=0$. Thus, at this limit, the EK system behaves like the LDM one: without ion current ($g_m=0$), the membrane is stable. Increasing $\beta$ represents increasing the effect of the membrane capacitance $C_m$, i.e. the ability to store charge at unit potential ($Q_m/V_m$), with $Q_m=|Q_i|$ (according to Eq.~\eqref{eq:macroscopic charge}) being the macroscopic charge along the membrane on one side. As $\beta \rightarrow +\infty $, $V_m=0$, and the $Q_m$ also reaches its maximum as $\gamma\rho_m$, with $\rho_m|_{\beta \rightarrow +\infty} =2V/(1+\gamma^2)$. The membrane capacitance $C_m=\varepsilon_m/d_0\rightarrow +\infty$ can be achieved by considering a membrane with zero-thickness $d_0\rightarrow 0$ but with finite $\varepsilon_m$. 
From an electrical standpoint, the membrane acts as a simple fluid–fluid interface: without potential jump across the membrane, the Debye layers on both sides still stores charges, and the internal normal traction from internal membrane disappears: $\rho_m^2\gamma^2/\beta|_{\beta \rightarrow 0}=0$.
In this limit, the Robin \zxy{condition} in Eq.~\eqref{eq:robin without dimension} reduces to the continuity of potential $\ljump\phi\rjump=0$, and the continuity of the displacement field $\ljump\varepsilon_f \bm{n \cdot E}\rjump=0$ at the interface. These two boundary conditions were used in an EK model for the simple fluid interface~\cite{Thaokar14_pre} that is impermeable to ions. It differs from the interfacial condition using ion conservation~\cite{Yariv15_jfm,Mori18_jfm}.
Mechanically, the membrane is different from a simple interface, due to the surface-incompressibility resulting from the conservation of lipids. The limit at $g_m\rightarrow0$ is also discussed in Ref.~\cite{Seiwert12_jfm}, which represents a short-circuited capacitor and the membrane acts as a simple fluid–fluid interface.

In Ref.~\cite{Seiwert12_jfm}, the growth rate monotonically increases with $g_m$ and then saturates, while in the current system, the growth rate first increases then decreases with $\beta$ and saturates to constant. This is due to the combination effect of internal and external effects. When excluding the internal normal traction (blue curve in Fig.~\ref{fig:beta effect}(a)), the growth rate behaves monotonically. In Fig.~\ref{fig:beta effect}(b), the external contribution to the growth rate increases with $\beta$ and saturates at a constant value, whereas the internal traction contribution initially rises from zero at $\beta = 0$, peaks, and then decreases back to zero as $\beta \to \infty$. Specifically, $\lim_{\beta\to 0}s_{\text{in}}=\lim_{\beta\to\infty}s_{\text{in}}=0$, corresponding to the overlap between the full growth rate (black curve) and the growth rate excluding internal traction (blue curve) at these limits in Fig.~\ref{fig:beta effect}(a). The internal traction contribution reaches its maximum at $\beta=\frac{\gamma}{1+\gamma}$, with $\max s_{\text{in}}=\frac{V^2\gamma^3}{1+\gamma}$. This peak results in a nonmonotonic behavior of the full growth rate in Fig.~\ref{fig:beta effect}(a), which is further highlighted by the variation of the maximum growth rate shown in Fig.~\ref{fig:beta effect}(c).

\begin{figure}[ht!]
    \centering
\includegraphics[keepaspectratio=true,width=0.9\columnwidth]{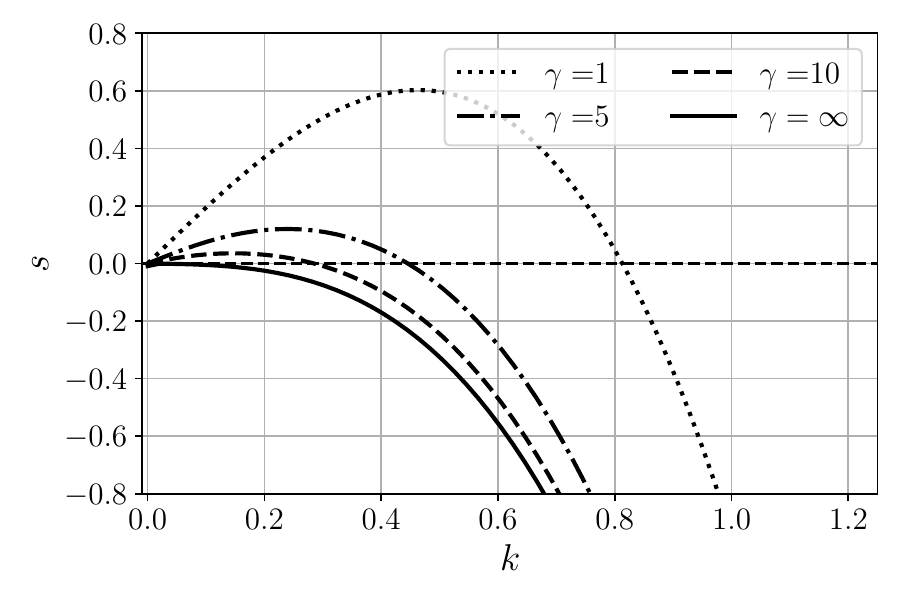}\put(-180,130){(a)}

\includegraphics[keepaspectratio=true,width=0.9\columnwidth]{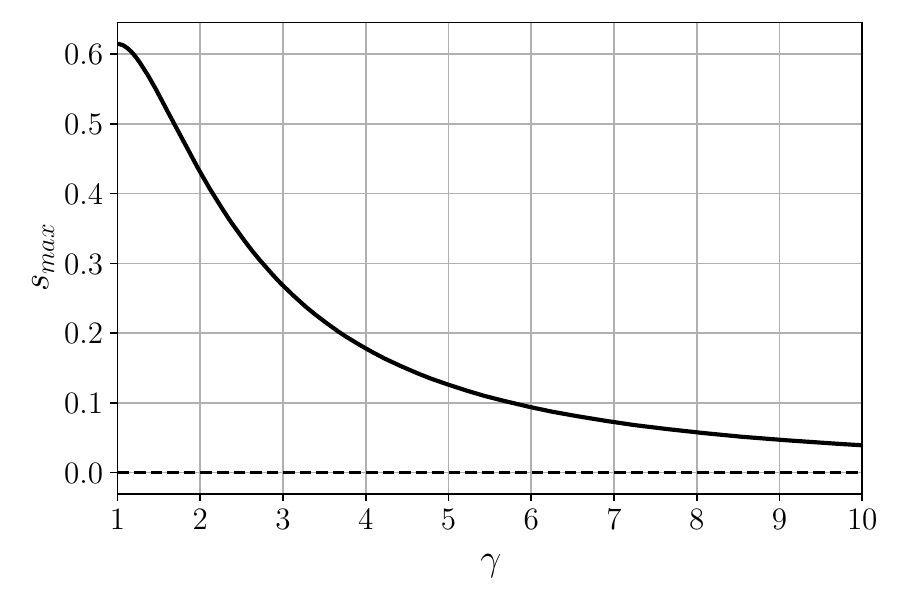}\put(-30,130){(b)}
    \caption{(a) The effect of $\gamma$ on growth rate $s$ as a function of perturbation wavenumber $k$; (b) maximal growth rate $s_{\mathrm{max}}$ as function of $\gamma$. We consider the system with \zx{$V=3$, $\beta=1$, $Ca=0.2$ and $\bar\Gamma=0.01$.}
    }
    \label{fig:R effect}
\end{figure}
Fig.~\ref{fig:R effect}(a) shows the effect of the asymmetric conductivity $\gamma$ on the growth rate. Due to the reversibility of the system, i.e. $s(\gamma)=s(1/\gamma)$, we only show results for $\gamma>1$. It is very counterintuitive that the membrane subjected to the fluids with the symmetric conductivity ($\gamma=1$) undergoes the most intense instability. As $\gamma$ increases from $1$, the maximum growth rate \zxy{decreases} as shown in Fig.~\ref{fig:R effect}(b), and the window for the unstable wavenumbers get narrower. As $\gamma\rightarrow+\infty$, the system becomes stable. This result is very different from the known literature~\cite{Seiwert12_jfm}, \zxy{where the instability is amplified as the conductivity becomes more asymmetric.} To understand the cause of this difference, it is essential to first examine the mechanism of the instability and how it diverges from the LDM framework. This will be explored in detail in the remainder of this section.

\subsection{Effect of electrostatic forces}
\label{sec:effect of forces}
\subsubsection{Stabilizing electrostatic body forces}
The Coulomb body forces in the Stokes equation~\eqref{eq:scaled stokes}, $\bm f=-\rho\nabla\phi$, tend to move the positive charges $\rho$ in the direction of the electric field $\bm E$. In the base state, these body forces are balanced by the pressure gradient on both sides of the fluid, as evidenced by Eq.~\eqref{eq:pressure p0}. Additionally, the flat interface remains free from normal traction arising from either electrostatic stress or elasticity, as well as membrane tension.
\begin{figure*}[ht!]
    \centering
\includegraphics[keepaspectratio=true,width=0.99\textwidth]
{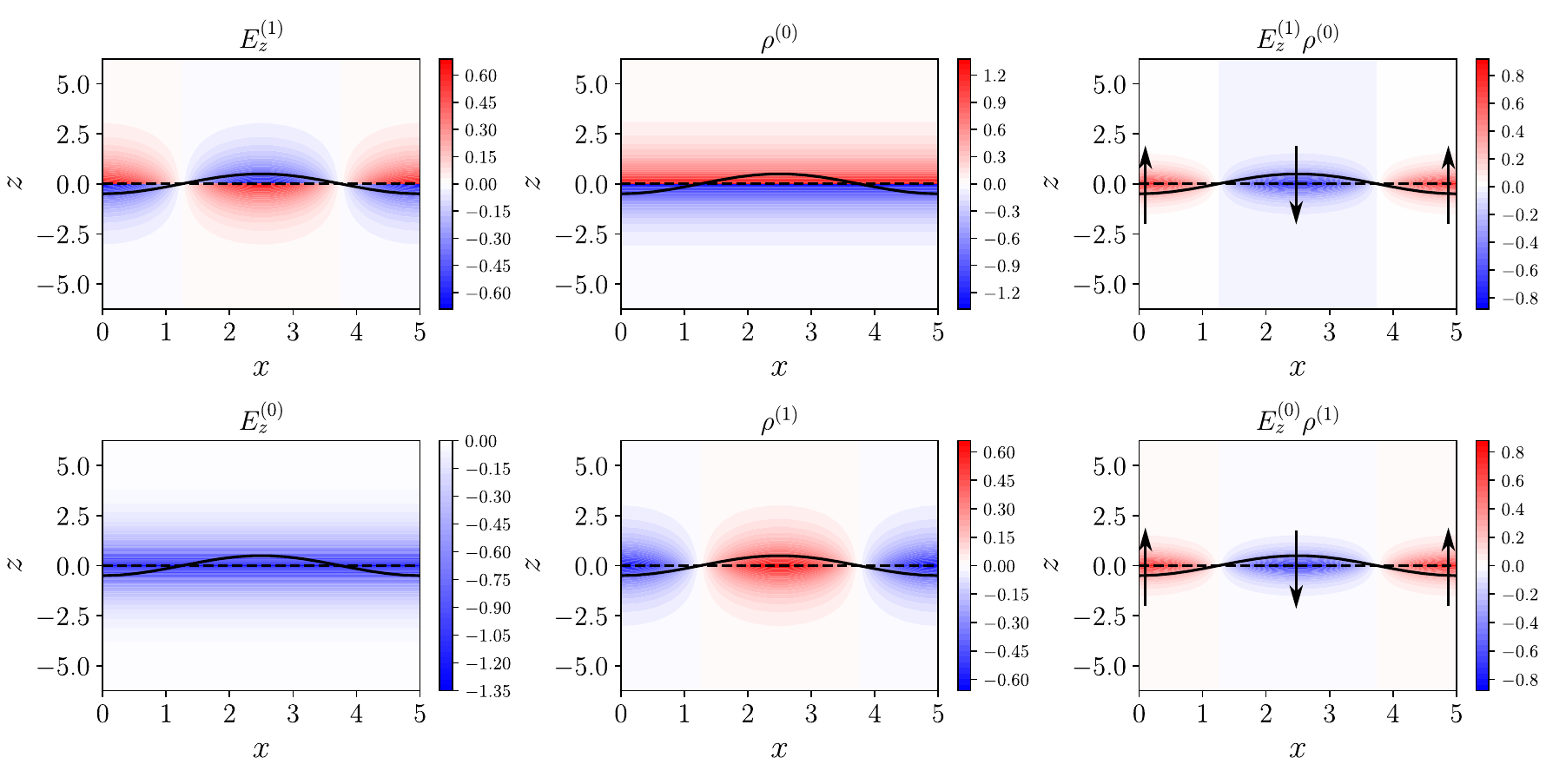}\put(-485,240){(a)}\put(-320,240){(b)}\put(-160,240){(c)}\put(-485,115){(d)}\put(-320,115){(e)}\put(-160,115){(f)}
    \caption{\zx{$z$-component of the electric field $E_z^{(0)}$ (a), $E_z^{(1)}$ (d) and charge density $\rho^{(0)}$ (b), $\rho^{(1)}$ (e), illustrate the components of the first-order electrical body force $\bm{f}^{(1)} \cdot \hat{e}_z = \rho^{(0)} E_z^{(1)} + \rho^{(1)} E_z^{(0)}$ shown in (c,f). The solid curve denotes the perturbed membrane, while the dashed line represents the flat membrane in the base state. Arrows in (c,f) indicate the direction of the $z$-component of the electrical body force. Notably, the force points downward at the crest and upward at the trough of the membrane perturbation, highlighting the restoring nature of the electrical body force. We consider symmetric conductivity $\gamma=1$, $\beta=1$, $V=1$.}
    }
    \label{fig:body force}
\end{figure*}

To understand the stabilizing terms in Eq.~\eqref{eq:growth rate full} and Eq.~\eqref{eq:body force expansion}, \zx{we illustrate the first order body force $\bm f^{(1)}=-(\rho^{(0)}\nabla\phi^{(1)}+\rho^{(1)}\nabla\phi^{(0)})$ in Fig.~\ref{fig:body force}(c,f), the 
$z$-component of the electric field $E_z^{(0)}$, $E_z^{(1)}$ in Fig.~\ref{fig:body force}(a,d), and charge density $\rho^{(0)}$, $\rho^{(1)}$ in Fig.~\ref{fig:body force}(b,e).} The \zx{electrical body force} is calculated with a uni-modal perturbation for wavenumber $k=0.2$, corresponding to a wavelength five times the Debye length, i.e., $x=5$. The perturbation amplitude is set to $h=0.5 $. It is expected that the \zx{electrical body force} acts only on the fluid in the Debye layer, decaying exponentially in the bulk. \zx{Fig.~\ref{fig:body force}(c,f) shows that within the Debye layer, the electrical body force points downward at the interface's crest and upward at the trough.} This implies that the electric body force plays a restoring role, consistent with the negative sign in the term Eq.~\eqref{eq:body force expansion} for small $k$.

\subsubsection{Destabilizing external electrostatic surface traction}

The external normal traction \zx{from the Maxwell stress} is
\begin{equation}
\ljump\partial_z\phi^{(0)}\partial_z\phi^{(1)}\rjump_{z=0}=
    (\partial_z\phi_{1}^{(0)}\partial_z\phi_{1}^{(1)}-\partial_z\phi_{2}^{(0)}\partial_z\phi_{2}^{(1)})|_{z=0}.
    \label{eq:surface traction general}
\end{equation}
In LDM framework~\cite{Seiwert12_jfm}, when the membrane is fully charged, and no current crosses the membrane, resulting in zero Ohmic current in the bulk fluid, the normal traction is zero due to the absence of a base state electric field $ \partial_z\phi^{(0)}_i=-E_z^{(0)}=0$.
In the current study, the base state electric field is nonzero at the membrane due to the Debye layer $ \partial_z\phi_{1}^{(0)}|_{z=0}=\partial_z\phi_{2}^{(0)}|_{z=0}=\gamma\rho_{m}$. Thus, Eq.~\eqref{eq:surface traction general} is further simplified as
 \begin{eqnarray}
&&\ljump\partial_z\phi^{(0)}\partial_z\phi^{(1)}\rjump_{z=0}\nonumber\\
&=&\partial_z\phi_{1}^{(0)}|_{z=0} \ljump\partial_z\phi^{(1)}\rjump_{z=0}
    = \partial_z\phi_{1}^{(0)} h\ljump\rho^{(0)}\rjump_{z=0}\nonumber\\
     &=&  \underbrace{-\partial_z \phi_{1}^{(0)}}_{\text{base state field}}
      (-h)
     \underbrace{\rho_{m}(1+\gamma^2)}_{\text{net charge density across membrane}}
     \label{eq:surface traction expain}
  \end{eqnarray}  
where $ \ljump\partial_z\phi^{(1)}\rjump_{z=0}=-h\ljump \partial_{zz}\phi^{(0)}\rjump_{z=0}=h\ljump\rho^{(0)}\rjump_{z=0}$ has been used according to the first order Robin \zxy{condition} in Eq.~\eqref{eq:Robin BC first}. Consequently, we interpret the external electrostatic surface traction as the net force exerted on the fluid adjacent to the membrane, namely the net charge density $\ljump\rho^{(0)}\rjump_{z=0}$ moving in the direction of the base state electric field $-\partial_z \phi_{1}^{(0)}$ (pointing downward), and modulated by the perturbed membrane at $(-h)$. After transforming from Fourier domain to physical domain, we find that the traction has the same formulation but with $h=h(x)$. Then, when $h(x)<0$, the surface traction points downward and vice versa, which renders the membrane unstable to deformations.  This implies the destabilizing effect of the external electrostatic normal traction, consistent with the positive sign in the term Eq.~\eqref{eq:growth rate full} for any small $k$. 

In addition, we observe that in contrast to the LDM, the presence of a non-zero base state electric field $\bm E^{(0)}$ primarily contributes to the instability. In the LDM, a non-zero base state electric field necessitates either bulk current~\cite{Seiwert12_jfm,Sens02_prl} or membrane charging. However, within the EK framework, $\bm E^{(0)}$ exists independently of other effects due to the Debye layer.
\begin{figure}[ht!]
    \centering
\includegraphics[keepaspectratio=true,width=\columnwidth]{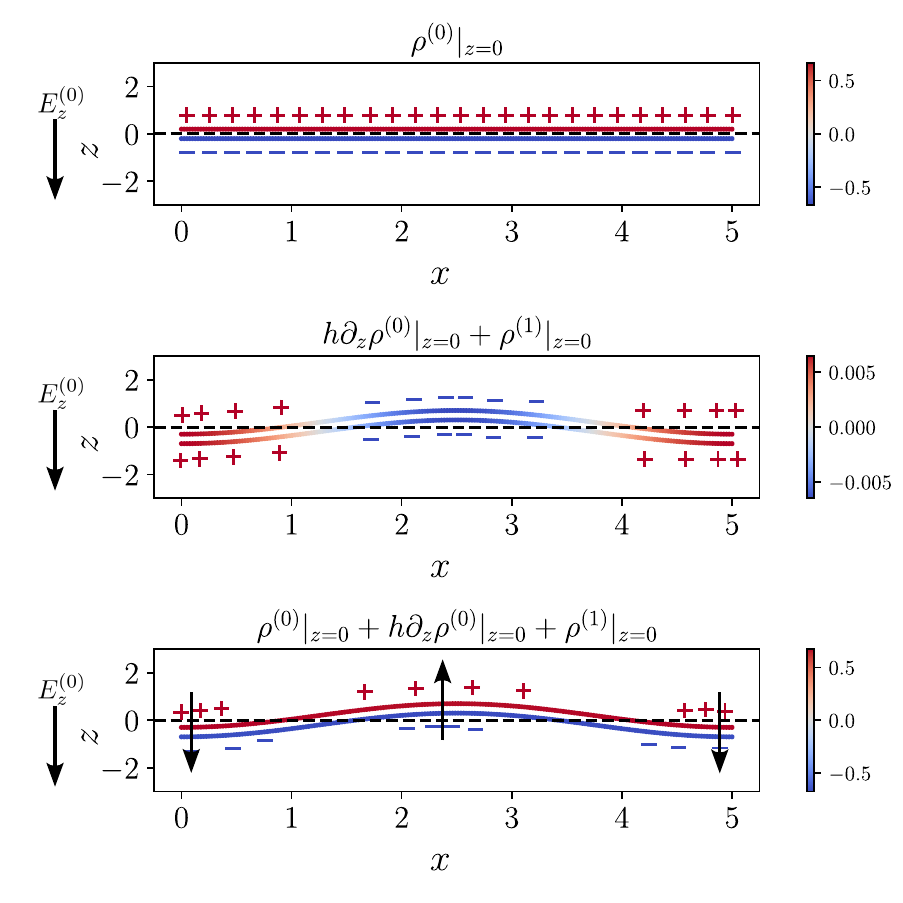}
\put(-240,230){(a)}\put(-240,150){(b)} \put(-240,70){(c)}
    \caption{\zx{(a) The uniform charge density in the base state. (b) The first-order perturbed charge density. The total charge density at the membrane, obtained by summing (a) and (b), is shown in (c). Membrane undulations induce a redistribution of charge along the interface, as illustrated in (c). Under a downward electric field, the resulting Maxwell stress exerts upward traction at the crests and downward traction at the troughs of the membrane, indicating its destabilizing effect on membrane shape. The figure shows the example with $\gamma=1$, $\beta=1$, $V=1$.
    }} 
    \label{fig:surface charge}
\end{figure}
Intuitively, we can understand such instability through the charge redistribution. Fig.~\ref{fig:surface charge} shows the charge distribution in the fluid along the membrane. At the base state (Fig.~\ref{fig:surface charge}(a)), the charge is uniformly distributed along the membrane. However, when subjected to a small perturbation (Fig.~\ref{fig:surface charge}(b)), the linear correction causes positive charge accumulation at the troughs on both sides of the membrane, while negative charge accumulates at the crests. As a result (Fig.\ref{fig:surface charge}(c)), on the upper surface of the membrane, positive charge accumulates denser at the troughs compared to the crests, while on the lower surface of the membrane, negative charge accumulates less at the troughs than at the crests. This accumulation further results in a force which tends to enhance the perturbation. Similar mechanisms of the instability resulting from surface charge redistribution due to membrane undulation are also discussed in Ref.~\cite{kumaran2000_prl}. Different from the present study, the surface charge in Ref.~\cite{kumaran2000_prl} are adsorbed on the interface, and such that compensates the charge distributed in the Debye layer, i.e. each lipid layer is electroneutral. 

\begin{figure*}[ht!]
    \centering
\includegraphics[keepaspectratio=true,width=0.99\textwidth]{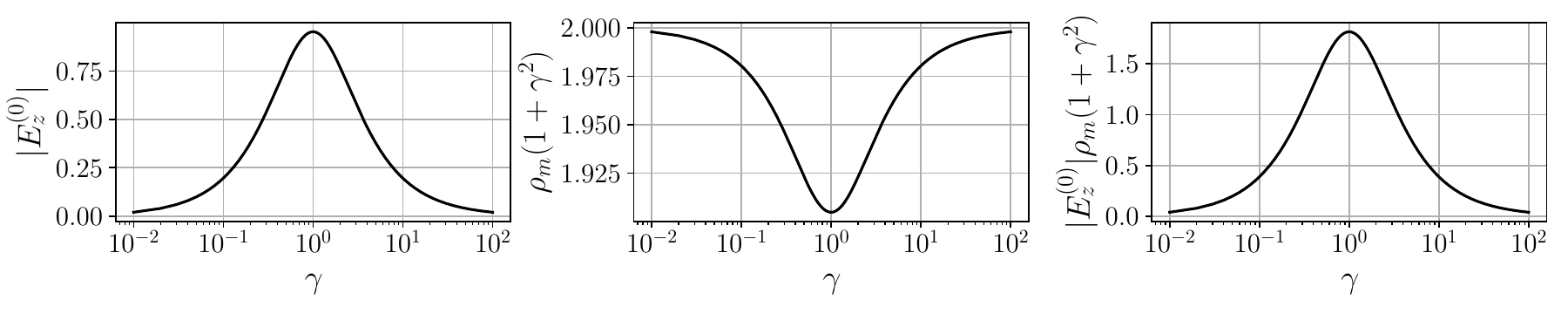}\put(-455,80){(a)}\put(-290,80){(b)}\put(-130,80){(c)}
    \caption{The effect of the conductivity ratio $\gamma$ on (a) base state electric field $E_z^{(0)}|_{z=0}$, (b) the base state charge density across the membrane, and (c) the external normal surface traction calculated according to Eq.~\eqref{eq:surface traction expain}. The example is given with $\beta=10$, $V=1$.
}
    \label{fig:asym-traction}
\end{figure*}

Next, we aim to understand \zxy{why the instability is reduced as the conductivity becomes more asymmetric.} We consider the scenario where $\gamma \rightarrow 0$, i.e. $R =\frac{n_1^\infty}{n_2^\infty}\rightarrow \infty$. 
In this limit, $\rho^{(0)}_1|_{z=0}=\rho_{m}=\frac{2\beta V}{\gamma+(1+\gamma^2)\beta}\rightarrow 2V$, $\rho^{(0)}_2|_{z=0}=-\gamma^2\rho_{m}\rightarrow 0$, and the net charge density across the membrane $\ljump\rho^{(0)}\rjump_{z=0}\rightarrow 2V$. However, the base state field $ \partial_z\phi_{i}^{(0)}|_{z=0}=\gamma\rho_{m}\rightarrow 0$, resulting in the entire external traction term approaching zero. Therefore, the instability decreases as $\gamma \rightarrow 0$. Similar results can be derived as $\gamma \rightarrow \infty$, where $\ljump\rho^{(0)}\rjump_{z=0}\rightarrow 2V$ and the base state field $ \partial_z\phi_{i}|_{z=0}=\gamma\rho_{m}\rightarrow 0$. The result of these two limits is illustrated in Fig.~\ref{fig:asym-traction}, where 
the destabilizing external surface traction attains its maximum when subjected to the symmetric conductivity.

\subsubsection{Destabilizing internal electrostatic surface traction}
The electric stress induced by the rearranged charge at the membrane has also been used to explain the internal normal traction in Ref.~\cite{Sens02_prl}, which establishes the system in the LDM framework, and the surface charge is brought to the interface by the bulk current. In their work, the compressive electrostatic stress has the form $\varepsilon_mE_m^2/2$. For a perfectly flat membrane, the electric stress in the membrane is symmetrically balanced on both sides of the membrane. When subjected to a perturbation, the redistributed charge $\Sigma_\pm\sim \pm \varepsilon_m E_m$ results into an unbalanced stress $(\Sigma_+^2-\Sigma_-^2)/\varepsilon_m$. This analysis shares the same situation as ours, where the base state macroscopic charge is $Q_i^{(0)}=\pm E_z^{(0)}|_{z=0}=\pm \beta d E_m^{(0)}$. 

In our work, the first-order internal traction has the form
\begin{eqnarray}
       - \bigg[\!\!\bigg[
\beta d\partial_z\phi_m^{(0)}\partial_z\phi_m^{(1)}
       \bigg]\!\!\bigg] _{z=\pm d/2}
  &=&
\beta d E_m^{(0)}
 \bigg[\!\!\bigg[
\partial_z\phi_m^{(1)}
       \bigg]\!\!\bigg] _{z=\pm d/2},
\end{eqnarray}
where the jump of the first order internal field $ \bigg[\!\!\bigg[\partial_z\phi_m^{(1)}\bigg]\!\!\bigg] _{z=\pm d/2}\sim
E_m^{(0)} dk^2h$ 
is nonzero due to the perturbation (see Appendix~\ref{app:internal normal traction} for expression). Intuitively, we can understand this jump as the virtual charges inside the membrane and driven by the field $E_m^{(0)}$. Like the external traction, we consider the deformation in physical domain, which has the same formulation but with $h=h(x)$. At the membrane where $h(x)<0$, the virtual charges are positive and driven by a negative internal field. Thus, the internal surface traction points downward and vice versa, which renders the membrane unstable to deformations. 
    
As discussed around Eq.~\eqref{eq:growth rate full}, the internal normal traction in this work is in agreement with the total traction in Ref.~\cite{Sens02_prl}, where the external traction vanishes in the limit of large $\varepsilon_m$ and small membrane conductivity $\sigma_m$. In their work, the internal field $E_m$ is related to the external field $E$ though the conservation of current, and thus $E_m\sim E\sigma_i/\sigma_m $, and thus small $\sigma_m$ indicates a large internal field. In our work, the internal field $E_m^{(0)}\sim E_z^{(0)}|_{z=0}/(\beta d)$, and thus $\beta d\ll 1$ indicates a large fields ratio $E_m^{(0)}/E_z|_{z=0}$. Keeping $d$ finite and letting $\beta \rightarrow 0$, we see the destabilizing effect of the normal traction dominated by the internal traction in Eq.~\eqref{eq:growth rate long}. This conclusion is in agreement with Refs.~\cite{Lacoste09_epje,lacoste07_el}, where the “inside” contribution is in general dominant when large voltage drop occurs across the membrane, and in our work $ \beta \rightarrow 0$ yields the largest transmembrane potential $V_m$ from Eq.~\eqref{eq:transmembrane potential}.

Similar to the external surface traction, as $\gamma\rightarrow 0$ or $\gamma\rightarrow +\infty$, the base state external field $E_z^{(0)}|_{z=0}\rightarrow 0$ and thus $E_m^{(0)}\rightarrow 0$, and the internal traction term vanishes.

\subsection{Spontaneous curvature}
\label{sec:curvature}
Biomimetic and biological membranes consist of two monolayers which are typically exposed to different environments and may differ in 
packing density of the lipid molecule's head and tail~\cite{Dobereiner00_is}. This asymmetry results in a torque across the membrane and such that the membranes prefer to curve and are characterized by the spontaneous curvature. In this study, the asymmetric conductivity of the electrolytes on both sides of the membrane changes the stress distribution, and we examine the spontaneous curvature $C_0$ calculated as the first moment of the lateral stress tensor~\cite{Loubet13_pre}. After scaling $C_0$ with $\lambda_D^{-1}$, we have
\begin{equation}
 C_0=Ca\int^{+\infty}_{-\infty}T_{xx}^{(0)} z dz,
\end{equation}
The component $T_{xx}$ is obtained as
\begin{eqnarray}
    T_{xx}=-p^{(0)}-\frac{1}{2}E_z^{(0)2}=-E_z^{(0)2},
\end{eqnarray}
where the base state pressure $p^{(0)}$ in Eq.~\eqref{eq:pressure p0} can also be obtained from $T_{zz}=-p^{(0)}+\frac{1}{2}E_z^{(0)2}=0$.
\zxy{Based} on this stress profile, the spontaneous curvature is obtained as
\begin{multline}
     C_0=Ca\int^{+\infty}_{-\infty}T_{xx}^{(0)} z dz =\frac{Ca \rho_m^2}{4}(1- \gamma^4).\\
     =Ca\beta^2 V ^2\frac{1- \gamma^4}{
     \left[\gamma+(1+\gamma^2)\beta\right]^2
     }.
\end{multline}
If $\gamma=1$, the spontaneous curvature vanishes in a symmetric system, and this corresponds to the result in~\cite{Loubet13_pre} for the finite-thickness membrane with zero charge density inside. However, the asymmetric conductivities of the bulk fluids tend to curve the flat membrane through the first moment of the stress. $\gamma>1$ corresponds to the conductivity $\sigma_1< \sigma_2$, and the membrane tends to have a negative curvature. The cell geometry altered by the asymmetric electric double layers adjacent to the membrane is also reported in experimental work~\cite{Faizi21_Electro}.

\zx{To illustrate the effect of spontaneous curvature and maintain consistency with earlier figures, we adopt the parameter set $\beta = 1$, $V = 1$, $\gamma = 2$, and $Ca = 0.2$. These values fall within the range typically considered in biologically relevant systems. With this choice, $\rho_m \approx 0.3$, the resulting spontaneous curvature is $|C_0| \approx 0.06$. Since $C_0$ increases with $\gamma$ for $\gamma > 1$, the value $\gamma = 2$ is selected to provide a representative example with non-zero curvature. In the limit $\gamma \to \infty$, the maximum spontaneous curvature reaches $|C_0|_{\max} = Ca V^2 = 0.2$. The $|C_0|\sim O(0.1)$ indicates that the spontaneous curvature is much smaller than the inverse Debye length. Thus, in the zero-thickness membrane assumption, we always have $|C_0|d \ll 1$, and it is valid to assume a flat membrane to conduct linear stability analysis~\cite{Seifert93_el}. Furthermore, the lateral membrane tension provides a stabilizing force that suppresses out-of-plane deformations. Thus, the influence of spontaneous curvature becomes subleading and does not qualitatively alter the results of the linear stability analysis.}

\subsection{Small perturbation assumption}
The infinitesimal perturbation assumption serves as a fundamental principle in the linear stability analysis of membranes or interfaces, a concept employed in EK modeling as seen in our present study and Ref.~\cite{Ziebert10_pre}, as well as in LDM, as demonstrated in Ref.~\cite{Seiwert12_jfm}. However, it becomes important to examine the perturbation amplitude relative to the Debye length.

The LDM assumes electroneutrality of the bulk fluid and free charge confined to the interface. Consequently, the model is inherently constrained by the thin Debye layer limit. Linear analysis derived from the LDM overlooks any dynamic behavior within the Debye layer, implicitly assuming perturbation amplitudes surpassing the Debye layer's dimensions.

In the EK model, when expanding the leading-order variables from $z=h$ to $z=0$, it becomes crucial to validate the linear approximation due to the anticipated abrupt change within the Debye layer. For instance, in the expansion of the Robin condition~\eqref{eq:robin without dimension}, we expand the base state electric field at the membrane as follows:
 \begin{equation}
\partial_z\phi_{i}^{(0)}|_{z=h}\approx \partial_z\phi_{i}^{(0)}|_{z=0}+h\partial_{zz}\phi_{i}^{(0)}|_{z=0},
\label{eq:expansion}
 \end{equation}
by assuming $h$ to be small.  However, when considering the exponential distribution of the base state potential $\phi^{(0)}_i$, we encounter $\partial_z\phi_{2}^{(0)}=Ae^{\gamma z}$. Substituting $z=h$ into $\partial_z\phi_{2}^{(0)}$, we find that to validate Eq.~\eqref{eq:expansion}, i.e.
\begin{equation}
    Ae^{\gamma h}\approx A(e^{\gamma 0} +h\gamma e^{\gamma 0})=A(1+\gamma h),
\end{equation}
the condition $|\gamma h| \ll 1$ is necessary. Similarly, after applying the same procedure to the upper fluid $\partial_z\phi_{1}^{(0)}$, we derive the full constraint $|h| \ll \min (\gamma,\gamma^{-1})$ on the perturbation amplitude to ensure the validity of the linear stability analysis. In cases where the conductivity on both sides of the membrane is symmetric, i.e. $\gamma=1$, and $|h| \ll 1$, we intuitively conclude that the linear analysis is only valid for perturbations smaller than the Debye layer.

Now, the discrepancy between the linear analyses stemming from the EK and LDM frameworks become apparent. In the EK model, the initial perturbation is assumed to be smaller than the Debye layer ($h\ll 1\ll L/\lambda_D$) and intensifies when exposed to instability induced by the electric field near the membrane. Conversely, the LDM presupposes perturbations larger than the Debye layer ($1\ll h\ll L/\lambda_D$), with the thin Debye layer riding atop the deformed membrane, which gets stabilized when subjected to a zero electric field in the bulk. This inherent distinction naturally prompts a fundamental question: How can we systematically determine the growth rate that effectively bridges the gap between the EK and LDM approaches, thereby accurately predicting the stability of perturbations with arbitrary amplitudes?" One potential solution is to explore the dynamics beyond the linear regime within the EK framework. By doing so, we anticipate observing the system transitioning to the base state, or other equilibrium states, through different manifolds when subjected to a significant perturbation.

An accompanying question arises: what is the final state of the unstable membrane in the current study, given that the perturbation cannot grow indefinitely due to the conservation of lipid molecules? We anticipate that nonlinearity will drive the system towards another equilibrium state, resulting in the emergence of stationary patterns in the membrane. For instance, Ref.~\cite{kumaran2000_prl} predicts a stable state with nonzero curvature as a consequence of the curvature–charge density coupling, where the transition from a flat membrane is induced when the Debye length is increased. This question can also be tackled by exploring nonlinearity, using methods such as fully nonlinear simulations or reduced nonlinear models like amplitude equations or long-wave equations as in Ref.~\cite{Young15_pof}. Extending the EK framework to incorporate nonlinear dynamics will be the focus of our future work.

\section{Conclusion and outlook}
\label{sec:conclusion}
The present work investigates the linear instability of a zero-thickness lipid membrane under a DC electric field, separating fluids with asymmetric conductivity. An electrokinetic model with the Debye-H\"{u}ckel approximation is adopted to capture the dynamics of the diffuse layers formed near the membrane. The capacitive effect of the membrane is found to be critical to the onset of the instability, due to which the charge accumulates inside the Debye layer. Upon perturbation, the charge redistributes and exerts a net force as surface traction on the fluid adjacent to the membrane, leading to destabilization. While the electrostatic body force tends to stabilize the system, it is unable to counteract the normal traction. The asymmetric conductivity stabilizes the membrane by reducing the base state electric field. This prediction diverges from that of the LDM framework, which overlooks the dynamics of charge within the Debye layer.

The developed analysis is applicable only to perturbations with amplitude smaller than the Debye layer. It is of interest to develop a model that systematically bridges the regimes of dimensional perturbation $h\ll\lambda_D$, $h\sim \lambda_D$ and $L\gg h\gg \lambda_D$. Such a model would be able to reconcile the contrasting stability predictions made by the present EK framework ($h\sim \lambda_D$) and the LDM framework ($L\gg h\gg \lambda_D$). 
One potential approach is to extend the current model into the nonlinear regime, allowing for illustration of the dynamical structure around the base state. Additionally, we anticipate that this nonlinear model will uncover other nontrivial equilibrium states, potentially revealing new patterns within the system.

Another intriguing factor to investigate is the non-equilibrium effect. The present study assumes instantaneous charge relaxation, and under this assumption, the Debye layer remains in equilibrium. However, in dynamic conditions, the bulk conduction process is inherently unsteady and influenced by ion convection due to fluid motion accompanying membrane deformation. In other words, with finite $\alpha$, the perturbation breaks the equilibrium state, causing charge in the bulk to move to or leave the Debye layer. This results in a transient current in the bulk that either charges or discharges the Debye layer. Such transient effects on stability have been studied in previous works~\cite{Schwalbe11_pof, Seiwert12_jfm, Ziebert10_pre}, where these effects introduce new instabilities in the system. However, in the EK framework, Ref.~\cite{Ziebert10_pre} neglects the ion convection and considers a constant current through the bulk and membrane that is independent of the perturbation in the linear order. It would be of interest to include finite $\alpha$ in the current system and investigate how membrane deformation and ion convection modulate instability. 

\zx{
This study explores the linear electrohydrodynamic response of a lipid membrane under a DC electric field through a minimal model designed to isolate the essential physical mechanisms. While motivated by the behavior of biological membranes, the model is not intended to reproduce the full complexity of living cells. Rather, it provides a simplified theoretical framework to gain insight into fundamental membrane-scale electrohydrodynamics. In this regard, artificial systems such as giant unilamellar vesicles (GUVs), which permit precise control over experimental conditions, offer a more directly relevant context for comparison~\cite{riske2005electro,Dimova09_sm}. For analytical tractability, the model omits several biophysical factors that may influence membrane behavior. In particular, it neglects in-plane lipid flows and intermonolayer friction, which can introduce dissipation and modulate the membrane’s dynamic response, especially in the nonlinear regime. Future extensions could incorporate a transport equation for lipid density~\cite{Schwalbe10_jfm} and include the corresponding dissipative forces~\cite{Evans1992dynamic} in the membrane force balance, enabling a more comprehensive description of membrane dynamics under electric fields.

}
\section{Acknowledgments}
This research was supported by NIGMS award 1R01GM140461.
\bibliography{reference}

\appendix
\section{Calculation of internal normal traction}
\label{app:internal normal traction}
The perturbed internal potential inside the membrane $z < |d/2|$ are governed by Laplace's equation
\begin{equation}
    (\partial_z^2-k^2)\phi_m^{(1)}=0,
\end{equation}
since there is no charge inside. The first-order internal potential is solved as
\begin{multline}
        \phi_m^{(1)}=\frac{ e^{\frac{d k}{2}}}{e^{2d k} - 1 }\left[
            \left(\phi_m^{(1)}|_{z=d/2} e^{d k} - \phi_m^{(1)}|_{z=-d/2}\right)e^{kz}\right.\\
    \left.+\left(\phi_m^{(1)}|_{-z=d/2} e^{d k} - \phi_m^{(1)}|_{z=d/2}\right)e^{-kz}
        \right].
         \label{eq:phi_m}
\end{multline}
The $\phi_m^{(1)}$ at the membrane is evaluated through the expansion of continuity of the potential
\begin{subequations}\begin{align}
    \phi_m^{(1)}|_{z=d/2}&=\phi_1^{(1)}|_{z=d/2}+(\pp{\phi_1^{(0)}}{z}|_{z=d/2}-\pp{\phi_m^{(0)}}{z}|_{z=d/2})h\nonumber\\
    &\approx[E_m^{(0)}+\rho_m  (\gamma-a_1)]h,\\
    \phi_m^{(1)}|_{z=-d/2}&=\phi_2^{(1)}|_{z=-d/2}+(\pp{\phi_2^{(0)}}{z}|_{z=-d/2}-\pp{\phi_m^{(0)}}{z}|_{z=-d/2})h\nonumber\\
    &\approx [E_m^{(0)}+\rho_m (\gamma-a_2)]h,
 \end{align}\label{eq:phi_m at membrane}
\end{subequations}
where external potential and field at the membrane are approximated using values at $z=0$. After substituting Eq.~\eqref{eq:phi_m at membrane} into Eq.~\eqref{eq:phi_m}, the internal normal stress is obtained as
\begin{eqnarray}
    &&    \bigg[\!\!\bigg[
 \frac{\varepsilon_m}{\varepsilon_f}\partial_z\phi_m^{(0)}\partial_z\phi_m^{(1)}
       \bigg]\!\!\bigg] _{z=\pm d/2}
  =
 \frac{\varepsilon_m}{\varepsilon_f}\partial_z\phi_m^{(0)} \bigg[\!\!\bigg[
\partial_z\phi_m^{(1)}
       \bigg]\!\!\bigg] _{z=\pm d/2}\nonumber\\
  &=& \beta d(-E_m^{(0)}) [2E_m^{(0)}+\rho_m  (2\gamma-a_1-a_2)]
\frac{k  \left(e^{d k} - 1\right)}{e^{d k} + 1}h\nonumber\\
  &=&-\frac{\rho_{m}^2\gamma^2}{\beta }[2-\frac{\beta d}{\gamma}(2\gamma-a_1-a_2)]\frac{k(e^{kd}-1)}{(e^{kd}+1)d}h
\end{eqnarray}
by substituting $E_m^{(0)}=-\rho_m\gamma/(\beta d)$.

In the limit $d\rightarrow 0$, the term $\frac{k(e^{kd}-1)}{(e^{kd}+1)d} \rightarrow k^2/2$.

With $d$ finite, and in the limit $k\rightarrow 0$, we evaluate 
\begin{eqnarray}
    (2\gamma-a_1-a_2)&=& \left[\gamma+\frac{\left(\gamma^2 - 1\right)^{2} }{2 \left(\beta \gamma^{2} + \beta + \gamma\right)}\right]   k^2+O(k^4),\nonumber\\
    \frac{k(e^{kd}-1)}{(e^{kd}+1)d}&=& \frac{1}{2}(k^2-\frac{d^2}{12}k^4)+O(k^6),\nonumber
\end{eqnarray}
and these terms are used to evaluate the growth rate at long-wave limit.

\end{document}